\documentclass[aps,prb,twocolumn,nopacs,floats,superscriptaddress]{revtex4-2}

\usepackage{graphicx}
\usepackage{dcolumn}
\usepackage{bm}
\usepackage{amsmath}
\usepackage{color}
\usepackage{amssymb}
\usepackage{ulem}
\usepackage{xcolor}
\usepackage{subfigure}

\usepackage{amssymb}
\usepackage{latexsym}
\usepackage{epsfig}

\usepackage{amsmath,amsfonts,amssymb,bm,ulem}
\usepackage{color}
\usepackage{float}

\definecolor{cardinal}{rgb}{0.6,0,0}
\definecolor{darkgreen}{rgb}{0,0.4,0}
\definecolor{golden}{rgb}{0.92, 0.7, 0}
\definecolor{midnight}{rgb}{0, 0, 0.5}
\definecolor{darkblue}{rgb}{0, 0, 0.7}

\def\he4{$^4$He}
\def\hel3{$^3$He}
\def\Am3{\AA$^{-3}$}
\def\beq{\begin{equation}}
\def\eeq{\end{equation}}

\newcommand{\be}{\begin{equation}}
\newcommand{\ee}{\end{equation}}
\newcommand{\bea}{\begin{eqnarray}}
\newcommand{\eea}{\end{eqnarray}}
\newcommand{\bse}{\begin{subequations}}
\newcommand{\ese}{\end{subequations}}

\def\Am2{\AA$^{-2}$}

\begin{document}

\title{Superclimbing modes in transverse quantum fluids: signature statistical and dynamical features}

\author{Chao Zhang}
\affiliation{Department of Physics, Anhui Normal University, Wuhu, Anhui 241000, China}

\author{Massimo Boninsegni}
\affiliation{Department of Physics, University of Alberta, Edmonton, Alberta, Canada T6G 2H5}

\author{Anatoly Kuklov}
\affiliation{Department of Physics \& Astronomy, College of Staten Island and the Graduate Center of
CUNY, Staten Island, NY 10314}

\author{Nikolay Prokof'ev}
\affiliation{Department of Physics, University of Massachusetts, Amherst, MA 01003, USA}

\author{Boris Svistunov}
\affiliation{Department of Physics, University of Massachusetts, Amherst, MA 01003, USA}
\affiliation{Wilczek Quantum Center, School of Physics and Astronomy and T. D. Lee Institute, Shanghai Jiao Tong University, Shanghai 200240, China}

%%%%%%%%%%%%%%%%%%%%%%%%%%%%%%%%%%%%%%%%%%%%%%%%%%%%%%%%%%%%%%%%%%%%%%%%%%%%%%
\begin{abstract}
Superclimbing modes are hallmark degrees of freedom of transverse quantum fluids
describing wide superfluid one-dimensional interfaces and/or edges with negligible 
Peierls barrier. We report the first direct numeric evidence of quantum shape fluctuations---caused by superclimbing modes---in simple lattice models, as well as at the free edge of an incomplete solid monolayer of $^4$He adsorbed on graphite. 
Our data unambiguously reveals the defining feature of the superclimbing modes---canonical conjugation of the edge displacement field to the field of superfluid phase---and its unexpected implication, i.e., that superfluid 
stiffness can be {\it inferred from density snapshots}.  
 
\end{abstract}

\maketitle
%%%%%%%%%%%%%%%%%%%%%
\section{Introduction}
%%%%%%%%%%%%%%%%%%%%%

The term ``superclimb" \cite{sclimb} emerged in the context of interpreting the effect of anomalous isochoric compressibility (a.k.a. syringe effect) accompanying superflow through solid observed in an imperfect \he4 crystal \cite{Hallock}. More specifically, it refers to the climbing motion of an edge dislocation supported by superfluidity of its core (distinguishing it from the conventional climb of edge dislocation supported by the pipe diffusion of vacancies along the core). From the very outset it was clear
\cite{sclimb} that if the Peierls barrier can be neglected  
the superclimb would have a profound impact on what otherwise would be standard superfluid phonon modes of a Luttinger liquid (LL).  The Hamiltonian $H[h,\phi]=\int dx {\cal H}$ of the effective long-wave field theory for such superclimbing edge 
can be written as
\be
{\cal H} \, =\, {\chi \over 2} (\partial_x h)^2 \, +\, {n_s \over 2} (\partial_x \phi)^2 \, ,
\label{H_TQF}
\ee
in terms of the canonically conjugate fields---the ``vertical" coordinate (height) of the edge $h(x)$, and the superfluid phase, $\phi(x)$, with $x$ the position along the edge. Here $n_s$ is the superfluid stiffness and $\chi$ is the edge tension. The dependence of the Hamiltonian density $\cal H$ on $\partial_x h$ rather than $h$ reflects translation invariance of $H$ with respect to the vertical motion of the edge as a whole, $h(x) \to h(x) + h_0$. To be more precise, we are dealing with 
discrete translation symmetry, but if the edge width, $d$, is significantly larger than the lattice distance, $a$, i.e. the edge is microscopically rough, 
then the Peierls barrier can be neglected on exponentially large
(larger than system size) length-scales.

By integrating out one of the canonically conjugate fields we obtain two equivalent actions  each suitable for straightforward computation of the remaining field properties. Starting from 
\be
S[h,\phi]\, =\,    \int (i h \, \partial_\tau \phi \, +\,  {\cal H}) \, dx d\tau  \, ,
\label{S}
\ee
we obtain
$S_{h}[h]$ and $S_{\phi}[\phi]$ from Gaussian integrals
\be
e^{-S_{h}[h]} = \!\int \!\! e^{-S[h,\phi]} \, {\cal D} \phi \, , \quad e^{-S_{\phi}[\phi]}  = \! \int \!\! e^{-S[h,\phi]} \, {\cal D} h  \, .
\label{actions}
\ee
In the Fourier representation, we have
\be
S_h = { 1 \over 2}\sum_{\omega, k}   \left[n_s^{-1}\, {\omega^2 \over k^2}
  \, + \, \chi k^2 \right] |h_{\omega,k}|^2 \, ,
\label{S_eta}
\ee
\be
S_\phi = { 1 \over 2}\sum_{\omega, k}  \left[\chi^{-1}\, {\omega^2 \over k^2}
  \, + \, n_s k^2 \right] |\phi_{\omega,k}|^2 \, .
\label{S_phi}
\ee
In what follows, we assume that $h(x)$ is counted from the equilibrium height value, i.e. 
$h(\omega , k=0) =0$.
The superclimbing modes described by the action (\ref{S_eta}) have  quadratic dispersion \cite{sclimb}
\be
\omega_k = Dk^2 \, , \qquad D=\sqrt{n_s \chi} \, ,
\label{omega_k}
\ee
involving two quite different types of quasi-one-dimensional motion
along and perpendicular to the edge: oscillations of the mass current 
and the geometric shape, respectively.

Recently, counter-intuitive properties and a much broader physics context 
for considering the model (\ref{H_TQF}) have been revealed, 
which lead to the formulation of the transverse quantum fluid (TQF) paradigm \cite{Kuklov2022,Radzihovsky2023,Kuklov2024a,Kuklov2024b}. It has been realized that infinite compressibility is the key ingredient defining the TQFs along with their unusual properties: (i) the quadratic dispersion relation for normal modes (or even the absence thereof), (ii) off-diagonal long-range order (ODLRO) at $T = 0$, and (iii) exponential dependence of the phase slip probability on the inverse flow velocity. From conceptual point of view, the TQF state is a striking demonstration of conditional character of many dogmas associated with superfluidity, 
including the existence of low-energy elementary excitations, in general, and the ones obeying the Landau criterion of homogeneous superflow stability, in particular.

Depending on the nature of ``transverse" particle reservoirs that guarantee the infinite compressibility of the quasi-1D superfluid, TQFs are divided into two subgroups: (i) the ``ordinary" TQFs and (ii) incoherent  ones (iTQF) \cite{Kuklov2024a}. In the TQF case, the reservoirs are insulating (gaped) states, qualitatively similar to the half-plane of atoms associated with an edge dislocation in an insulating solid. 
Infinite effective 1D compressibility in such reservoirs is exclusively due to the translation invariant superclimbing motion of the edge. The gaped reservoir modes play no role in the long-wave physics, which, therefore, is universally described by the 1D Hamiltonian (\ref{H_TQF}).
In the iTQF case, the edge is not climbing and at least one of the reservoirs must be gapless
because its soft modes are required for infinite 1D compressibility and, thus, 
play crucial part in the long-wave dynamics of the superfluid phase. In some cases,
the resulting dynamics has a diffusive character and is described by an effective action $S_{\phi}$
rather than an effective Hamiltonian (see Ref.~\cite{Kuklov2024a} and references therein). While being quite different in terms of the linearized long-wave dynamics, TQF and iTQF both feature ODLRO and share similar instanton physics \cite{Kuklov2024a}.

The  fingerprints of TQFs are universal off-diagonal space--(imaginary-)time correlations. Despite
radical difference between the TQF and iTQF cases in terms of the underlying physics, 
their phase correlations are described by self-similar functions derived from the 
corresponding effective theories \cite{Radzihovsky2023,Kuklov2024a} and revealed 
in recent {\it ab initio}  simulations of simple iTQF models in Ref.~\cite{Kuklov2024b}.
In Matsubara representation, many aspects of the long-range and mesoscopic off-diagonal order in TQF and iTQF look similar.  
In particular, for both cases $S_\phi[\phi]$ is characterized by infinite compressibility, which in terms of the Fourier components of the field $\phi$ is defined as \cite{Kuklov2024a}
\be
\kappa\, =\,  \lim_{\omega \to 0} {1\over \omega^2} \lim_{k_x\to 0} {\delta^2 S_\phi [\phi_{\omega,k_x},\phi_{\omega,k_x} ^*] \over \delta \phi_{\omega,k_x}  \delta  \phi_{\omega,k_x}^*} \,=\, \infty \, .
\label{compress}
\ee
This property is both necessary and sufficient to ensure ODLRO and
power-law binding of instantons irrespective of other system details such as 
gaped or gapless bulk excitations, spectrum of elementary excitation, or their very
existence. At $\kappa \neq \infty$, the low-energy physics of the system would corresponds to that of LL.

While emphasizing deep similarities between TQF and iTQF, we should also keep in mind important qualitative differences. The key feature distinguishing TQF from iTQF is the presence of well-defined and most specific elementary excitations---the superclimbing modes, which are the main focus of this work.  
We present numerical evidence supported by rigorous quantitative analysis that edge/interface shape fluctuations in two distinct 2D lattice models are indeed 
originating from superclimbing TQF modes. 
The first model  describes phase separated one-component hard-core bosons 
(microscopically equivalent to an easy-axis $XY$ ferromagnet) while the second one deals 
with the two-component soft-core bosons at integer total filling in the regime of phase separation when each component is in the Mott-insultor (MI) phase. In both models, the bulk phases are insulating. In the hard-core system, the interface 
is a standard superfluid, while in the two-component case, the phase boundary is a supercounterfluid \cite{SCF}. Despite these microscopic differences, 
the emerging low-energy physics turns out to be the same and that of the TQF. 

In an attempt to identify an actual physical system in which this behavior could be observed experimentally, we carried out microscopic numerical simulations of an incomplete $^4$He monolayer adsorbed on graphite. Our results indicate that this system is a promising candidate for a TQF at the free edge.

%%%%%%%%%%%%%%%%%%%%%%%%%%%%%%%%
\section{Correlators}
%%%%%%%%%%%%%%%%%%%%%%%%%%%%%%%%

The simplest correlation functions revealing universal TQF fluctuations are \cite{Radzihovsky2023}
\be
\tilde K(x,\tau) \, =\, {1\over 2}\,  \langle \, [\, h(x,\tau)-h(0,0)\, ]^2 \, \rangle
\label{hight_corr}
\ee
and
\be
\tilde F(x,\tau) \, =\, {1\over 2}\,  \langle \, [\, \phi(x,\tau)-\phi(0,0)\, ]^2 \, \rangle \, .
\label{phase_corr}
\ee
Phase correlations control the asymptotic behavior of the correlator  of the superfluid order-parameter field
$\psi \propto e^{i\phi}$:
\[
\langle \, \psi(x,\tau)\,  \psi^*(0,0) \,  \rangle \, \propto \,
\left\langle \, e^{i[\phi(x,\tau)-\phi(0,0)]} \, \right\rangle\, = \, e^{-\tilde{F}(x,\tau)} \, . 
\]
In the low/zero-temperature regime, and at large $|x|$ and/or $|\tau|$, correlators (\ref{hight_corr}) and (\ref{phase_corr}) have the same universal functional form, 
if one makes a substitution
\be
n_s \, \leftrightarrow \, \chi 
\label{swap}
\ee
after subtracting non-universal additive constants $K_\infty$ and  $F_\infty$, see Eqs.~(\ref{hight_LRO}) and (\ref{phase_LRO}) below. This is the hallmark of TQF implied by the fact that the Hamiltonian (\ref{H_TQF}) is symmetric with respect to simultaneously swapping $h(x)$ with $\phi(x)$ and $\chi$ with $n_s$. Compactness of the field $\phi(x)$ is irrelevant for the asymptotic behavior. However, 
in a large but finite system with periodic boundary conditions and at a low but finite temperature, the compactness of the field $\phi(x)$ requires taking into account states with nonzero phase winding numbers (see in Ref.~\onlinecite{Kuklov2024b}). Such a generalization is absolutely straightforward  because fluctuations of the phase winding numbers are purely classical and statistically independent from all other fluctuations.
In this work, we focus on the height-height correlations and their precise connection to the compact phase-phase correlations. 

The prominent feature of (\ref{hight_corr}) and (\ref{phase_corr}) is the long-range order expressed as saturation of both correlators to finite values in the thermodynamic limit at $T=0$:
\be
K_\infty \, \equiv \, \tilde K(\infty, \infty)  \, =\,  \langle \, [\, h(0,0)\, ]^2 \, \rangle \, < \, \infty \, ,
\label{hight_LRO}
\ee
\be
 F_\infty   \, \equiv \, \tilde F (\infty, \infty) \, =\, \langle \, [\, \phi(0,0)\, ]^2 \, \rangle \, < \, \infty \, .
\label{phase_LRO}
\ee
Physically, Eq.~(\ref{hight_LRO}) means that the edge is asymptotically smooth (despite appearing 
quantum rough on the edge-width scale), while Eq.~(\ref{phase_LRO}) is the ORLRO statement 
implying Bose-Einstein condensation because the field correlator at infinity $\propto e^{- F_\infty} \neq 0$. For a number of fundamental and circumstantial reasons, the values of
$K_\infty$ and $F_\infty$ are nonuniversal, and, to a certain extent, arbitrary. 
Indeed, the harmonic form of the Hamiltonian (\ref{H_TQF}) implies a certain system-dependent UV cutoff for the fields  $h(x)$ and $\phi(x)$ beyond which the physics is neither harmonic nor universal. Furthermore, the fact that $h(x)$ and $\phi(x)$  are collective rather than microscopic 
variables implies that these do not have unique definitions at short distance. These ambiguities are removed by dealing with relative quantities:
\bea
K(x,\tau) \, &=&\, \tilde K(x,\tau)\, -\, K_\infty\, , \label{K_rel} \\
 F(x,\tau) \, &=&\, \tilde F(x,\tau) \,  -\,  F_\infty\, .
\label{C_rel}
\eea
At large enough $|x|$ and/or $|\tau|$, the behavior of $K(x,\tau)$ and $F(x,\tau)$ 
is universal and fully controlled by the effective harmonic action
(\ref{S_eta})--(\ref{S_phi}).

Since two actions are identical up to the substitution (\ref{swap}), in what follows we proceed with analysing one of them; specifically, we consider the phase action $S_\phi$ 
and evaluate (\ref{C_rel}). Despite our ultimate interest in its counterpart (\ref{K_rel}), 
we prefer to work with phase fluctuations thereby emphasizing close qualitative and quantitative similarities between the two. After mentioning ground-state thermodynamic limit results reported 
in Ref.~\onlinecite{Kuklov2024b} (for the purpose of self-contained presentation) we derive and test numerically analytic TQF predictions for 
finite-size system at non-zero temperature. 

%%%%%%%%%%%%%%%%%%%%%%%%%
\section{Ground-state fluctuations}
%%%%%%%%%%%%%%%%%%%%%%%%%%

We start by reminding that all results apply to both $K(x,\tau)$ and $F(x,\tau)$ because these functions are related to each other by the transformation (\ref{swap}): 
\be
\chi K(x,\tau)  =   n_s F(x,\tau)  \equiv  C(x,\tau) = - \! \int_{k,\omega} \!\! e^{ik x+ i\omega \tau}  c(\omega,k) ,
\label{Q}
\ee
\be
c(\omega,k)\, = \, \frac{k^2}{(\omega/D)^2 + k^4} \,  .
\label{TQF}
\ee
Here $\int_{k,\omega} \, \equiv \int \frac{d\omega dk}{(2\pi)^2}$.  Parameter $D$ is invariant under the transformation (\ref{swap})
and thus is the only TQF constant controlling the shape of correlation functions. To be precise, the integral over $k$ in (\ref{Q}) contains an ultra-violet cutoff, $k_0$, irrelevant at large enough $|x|$ and/or $|\tau|$ but setting limitations on the applicability of  (\ref{Q})--(\ref{TQF}) when both $|x|$ and/or $|\tau|$ are small: $|x|\lesssim k_0^{-1}$ and $|\tau| \lesssim (Dk_0)^{-2}$. For microscopically 
wide edges $k_0 \sim 1/d$. 

Straightforward integration over $\omega$ in (\ref{Q}) results  in
a Gaussian integral over $k$ and the final answer for the ground state in the thermodynamic limit:
\be
C (x,\tau)  \,=\, - {\sqrt{D} \, e^{-{x^2 \over 4 D |\tau |}} \over 4 \sqrt{\pi |\tau |}} \, .
\label{C_macro}
\ee
It is very instructive to compare this expression with its LL counterpart. While the latter is scale-invariant and isotropic (up to rescaling of the imaginary-time variable) in the $(1+1)$-dimensional Euclidean spacetime, the former is not. The prominent feature of Eq.~(\ref{C_macro}) is self-similarity: Up to a scale-invariant prefactor $\propto |\tau |^{-1/2}$, the dependence on $x$ and $\tau$ reduces to a function of the dimensionless argument $x^2  (D |\tau |)^{-1}$. Similar type of scaling (with different meaning of the parameter $D$) describes correlations in the iTQF state \cite{Kuklov2024b}.

A striking feature distinguishing TQF from iTQF is the form of the equal-time correlator $C (x,0)$. It is supposed to be a 
decaying power-law function of $x$, and consistent with this observation, in iTQF we have $C (x,0) \propto 1/|x|$ \cite{Kuklov2024b}. 
In a sharp contrast to that, the TQF case proves to be special because
\be
C (x,0)\equiv 0 \, , \qquad x\neq 0\, ,
\label{C_equiv_0}
\ee
meaning that equal-time universal correlations  are simply absent in the TQF ground state.
[It should be noted that this result does not apply to fast 
non-universal short-range decay governed by the UV physics.]
This is a remarkable example of how zero-point fluctuations above the UV limit can sum up into exact zero; which is important in the context of distinguishing TQF from both LL and iTQF. The result (\ref{C_equiv_0}) points to a certain subtlety when it comes to the problem of revealing universal quantum mechanics of the superclimbing modes experimentally
because equal-time correlators are the most natural experimental observables, e.g. for the ultracold-atomic systems. 
However, the situation changes significantly when non-zero 
temperature satisfies the condition $D\beta < L^2$, 
where $L$ is the edge length. 

%%%%%%%%%%%%%%%%%%%%%%%%%
\section{Non-zero temperature}
%%%%%%%%%%%%%%%%%%%%%%%%%%

Finite temperature is a resource that should be analyzed with care. 
On the one hand, superclimbing modes need to be excited to become visible in the universal signal $C (x,0)$. On the other hand, classical
harmonic behavior settles in at large distances because the dominant contribution is coming from modes with large occupation numbers, rendering correlations trivially universal. In the classical limit, the fields $h$ and $\phi$ become statistically independent,
so that their thermal fluctuations are independently controlled by the parameters $\chi$ and $n_s$, respectively. This, in particular,  means than all the individual features of TQF are lost in this limit.

To reveal the universal behavior at finite $T$ it is sufficient to subtract the {\it zero-temperature} constant $C_\infty$ from the finite-temperature correlator $\tilde{C}(x,\tau)$, i.e. there is no need to modify relations (\ref{C_rel}). 
At low but finite $T$ all we need is to replace frequency integrals
with Matsubara sums over $\omega_m=2\pi m T, \, m=0,\pm 1, \pm 2, ...,$ 
\be
\tilde{C}(x,\tau)= \sum_m \! \int^{k_0} \!\!\! {dk\over 2\pi \beta}\left[1-\! e^{i\omega_m\tau +ikx}\right] \! c(\omega_m,k)+C_*(k_0).
\label{Ctil}
\ee
Here we explicitly cut off the momentum integration at appropriately low momentum $k_0$ with simultaneously introducing cutoff-dependent (and temperature-independent) constant $C_*(k_0)$, which absorbs all non-universal UV contributions. With this parametrization, 
\be
C_\infty \, =\, \int  {d\omega \over 2\pi }\,\int^{k_0}\!  {dk\over 2\pi} \, c(\omega,k) +C_*(k_0) \, .
\label{Cinf}
\ee
This brings us to the following decomposition for $C(x,\tau) $ (in which we safely take the limit $k_0\to \infty$ assuming that $|x|$ and/or $\tau$ are large enough)
\be 
C(x,\tau) \, =\, C_0  \, +\, C_{\rm qu}(x, \tau) \, +\, C_{\rm cl}(x) \, ,
\label{C_finite_T}
\ee
where
\be
C_{\rm qu}(x, \tau) \, =\, - 
\sum_{m\neq 0} e^{i\omega_m \tau} \!\!\! \int \!\! {dk  \over 2\pi \beta}  \, e^{ik x} \, c(\omega_m,k) 
\label{C_qu} 
\ee
is the quantum contribution coming from nonzero Matsubara frequencies and vanishing at $|x|\to \infty$,
\be
C_{\rm cl}(x) \, =\, \int \!\! {dk  \over 2\pi \beta}  \, \left( 1 - e^{ik x}\right) c(0,k) 
\label{C_cl}
\ee
is the (diverging with $|x|$) classical contribution coming from 
zero Matsubara frequency, and
\be
C_0 =  \int {dk\over 2\pi}  \left[ T \sum_{m\neq 0}  \, c(\omega_m,k) \,  -\,  \int {d\omega \over 2\pi } c(\omega,k) \right]
\label{C_0_finite_T}
\ee
is a temperature-dependent universal constant [guaranteeing, in particular, that $C(x,\tau)$ approaches  the expression (\ref{C_macro}) in the limit $|\tau|\ll \beta$].

Performing straightforward integration over $k$ in Eqs.~(\ref{C_qu})--(\ref{C_0_finite_T}), we find
\be
C_0 \, =  - c_0  \sqrt{D \over \pi  \beta}\, ,
\label{C_0_finite_T_res}
\ee
where 
\be
 c_0 = \!\!\!\ \lim_{m_*\to \infty} \left[ \sqrt{m_*+1/2} \, -  \sum_{m=1}^{m_*} {1 \over 2 \sqrt{m}} \right]  \approx 0.730 \, ,
\label{c0}
\ee
\be
C_{\rm cl}(x) \, =\, {|x| \over 2\beta} \, ,
\label{C_cl_res}
\ee
and (using a compact notation $s=2\tau/\beta \in [0,2]$)
\be
C_{\rm qu}(x, \tau) \! =  \! \sqrt{D \over 2 \pi \beta} \sum_{m = 1}^{\infty}  {e^{-{\sqrt{m}x\over x_*}}\!  \cos (\pi m s)  \sin ( {\sqrt{m}x\over x_*} \! - \! {\pi \over 4} )  \over \sqrt{m}} ,
\label{C_qu_res}
\ee
\be
x_* \, =\, \sqrt{D\beta \over \pi} \, .
\label{x_*}
\ee
We need to take the limit in (\ref{c0}) because after 
integrating over momenta the remaining frequency integral and sum
in Eq.~(\ref{C_0_finite_T}) are separately UV divergent; 
the term $1/2$ under the square root dramatically enhances 
convergence to the limit.

Somewhat counter-intuitively, it is still possible to extract parameter $D$ from fluctuations at distances $|x| \gg x_*$, 
where the $x$-dependent quantum contribution (\ref{C_qu_res}) is exponentially suppressed and the classical contribution (\ref{C_cl_res}) increases linearly with $|x|$: the sub-leading 
constant term $C_0$ is controlled by $D$.

%%%%%%%%%%%%%%%%%%%%%%%%%
\section{Non-zero temperature and finite system size}
%%%%%%%%%%%%%%%%%%%%%%%%%%

The only difference between the $T\ne 0$ treatment of correlations 
in infinite and finite systems (with periodic boundary conditions) is coming from replacing momentum integrals with discrete sums over $k_n=2\pi n /L, \, n=0,\pm 1, \pm 2, ...,$ in 
Eqs.~(\ref{Ctil})--(\ref{C_0_finite_T}).
In this case, the universal part of the correlator, $C (x,\tau)$, approaches the ground-sate thermodynamic value (\ref{C_macro}), in the limit $|x|\ll L, \, |\tau|\ll \beta$;  it also reproduces the finite-temperature thermodynamic value (\ref{C_finite_T}) at $|x|\ll L$.
Using compact notation $r=2 x/L \in [0,2]$ we have
\be C(x,\tau) = C_0 - \!\!\!\!\!\!\!\! \!\!\!\! \!\!
\sum_{\scriptsize \begin{array}{c} m,n =-\infty\\(|m|+|n| \neq 0) \end{array}} ^{\infty}  
e^{i\pi n r + i \pi m s} \, c_{mn} \;, 
\label{Lambda_discr_2}
\ee
with 
\be
c_{mn} = \frac{c(\omega_m,k_n)}{\beta L} \equiv  {\pi^2 D^2 \beta \over  L^3}\,  {n^2 \over (\pi m)^2 + (\nu n^2)^2} \, ,
\ee
\be
\nu = {2\pi^2 D \beta\over L^2} \,, 
\label{nu-parameter}
\ee
and
\be
C_0 =    \sum_{\scriptsize \begin{array}{c} m,n =-\infty\\(|m|+|n| \neq 0) \end{array}} ^{\infty}   c_{mn}  \;
-  \int_{\omega,k} c(\omega,k) .
\label{Lambda_0}
\ee
In the last expression, both the integral and the sum have 
ultraviolet momentum cutoffs, which mutually cancel to the leading order.

After summation over $m$ using standard relation
\[
 \sum_{m=-\infty}^{\infty} \!{\cos( \pi m s ) \over (\pi m)^2 + a^2} = {\cosh[(1 - s) a]\over a \sinh a}
\]
we get
\be
C(x,\tau)  = {L \over 2 \beta }  \, \bar{\lambda}(r,s) \, ,
\label{two_lambdas}
\ee
\be
\bar{\lambda}(r,s) \, =\, \lambda_0 + \lambda_1(r,s)\, ,
\label{barlam}
\ee
\be
\lambda_0 =    {\nu\over \pi^2} \sum_{n=1}^{\infty} \left[  \coth(\nu n^2)  - 1 \right] \, ,
\label{lambda_0}
\ee
\be
\lambda_1(r,s)  =  - {\nu\over \pi^2}\! \sum_{n=1}^{\infty}  \cos(\pi n r)  {\cosh(|1- s| \nu n^2) \over \sinh(\nu n^2)}\, .
\label{lambda_1}
\ee
Special care should be taken of the important equal-time case 
$s=0$ [or $s=2$, which is the same by the symmetry of the expression (\ref{lambda_1})]. At $s=0$, the series (\ref{lambda_1}) does not converge. It is easy to see, however, that the problem is all about the $\delta$-functional contribution developing at $r=0, 2$  as  $s \to 0$. While remaining meaningful up to certain small non-zero
values of $s$, the diverging contribution ultimately becomes non-physical and thus needs to be removed from the answer for $\lambda_1(r,s)$ once its width gets smaller than the characteristic microscopic cutoff scale. From the theory of Fourier series we  have
\[
 {1\over 2}\delta(r) \, + \, {1\over 2}\delta(2-r)  \, =\, {1\over 2}\, +\, \sum_{n=1}^{\infty} \, \cos(\pi n r) \, .
\]
This observation provides us with counter-terms for taking the limit $\tau \to 0$ of the series (\ref{lambda_1}) at $0 < x <2$. We thus get
\be
\bar{\lambda}(r,0) =  {\nu\over \pi^2}\left\{ {1\over 2} + \sum_{n=1}^{\infty} \, [1\! -\! \cos(\pi n r)] \left[  \coth(\nu n^2)\!  - \!1 \right] \right\} .
\label{lambda_bar_0}
\ee

\section{Hard-core bosons: Mott-insulator--vacuum interface}

\begin{figure}[t]
\includegraphics[width=0.9 \columnwidth]{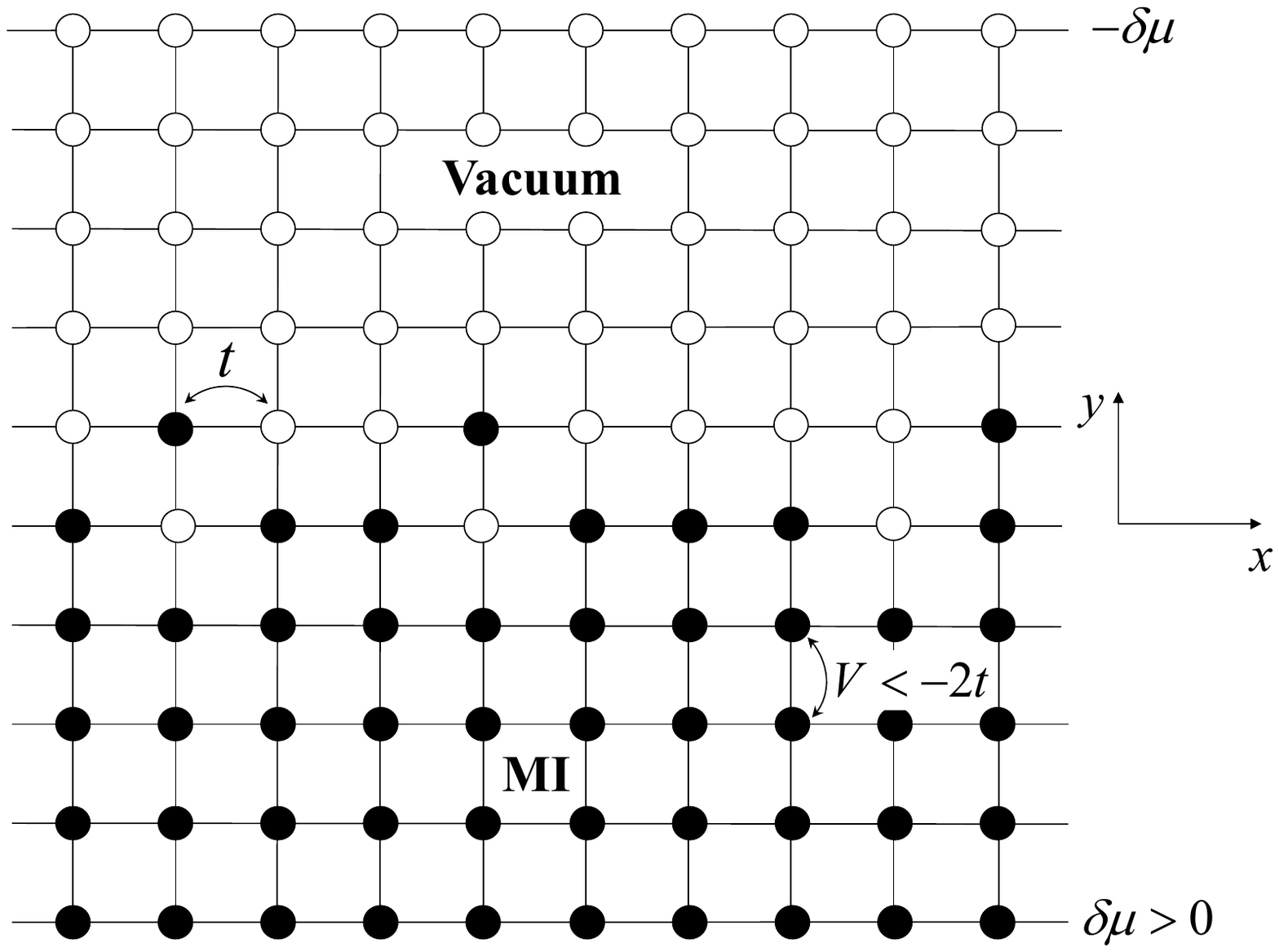}
\caption{Phase separated state of the hard-core bosonic Hamiltonian with nearest-neighbor
attraction $V<-2t$ at half filling. It can be also viewed as the phase separated state
of the easy-axis $XY$ ferromagnet at zero total magnetisation.}
\label{hard_core}
\end{figure}

The Hamiltonian of the hard-core system is arguably the most simple
of all possible TQF realizations. It consists of the nearest-neighbor
hopping and interaction terms on the square lattice:
\begin{equation}
H_{hc} = -t \sum_{\langle {i,j} \rangle } b_{j}^{\dagger} b_{i}^{\,}
        + V  \sum_{\langle {i,j} \rangle } n_{j} n_{i}    \;,
\label{Hhc}
\end{equation}
with the constraint on the occupation numbers, $n_{i}\le 1$ (here $b_{i}$ is the bosonic annihilation operator on site ${i}$).
In what follows, we use the hopping amplitude, $t$, and the lattice constant as units of energy and length, respectively.
The model can be re-written identically as the spin-$1/2$ ferromagnetic $XYZ$ model with
$J_x=J_y=-2t$ and $J_z=-V$. At $V=-2t$ particles gain as much energy, $-4t$, from attractive
interactions in the MI state with $n_i=1$ as they get from delocalization in an empty
lattice, or vacuum state, with $n_{i} \to 0$. By decreasing $V$ below $-2t$, we ensure a
MI-vacuum phase-separated state with the width of the interface diverging as $V \to -2t$.
To create such an interface oriented along the $x$-direction it is sufficient to pin the
structure by adding small potentials $\pm \delta \mu$ at the lattice edges in the $y$-direction.
This setup is illustrated in Fig.~\ref{hard_core} for half-filled lattice. What makes this model
very efficient numerically, is absence of quantum fluctuations in the bulk, i.e. the ``active" simulation volume is limited to the close vicinity of the domain wall.
%%%%%%%%%%%%%%%%%%%%%%%%%%%%%%%%%%%%%%%%%%%%%%%%%%
%%%%%%%%%%%%%%%%%%%%%%%%%%%%%%%%%%%%%%%%%%%%%%%%%%%%%%%

The model (\ref{Hhc}) has been simulated using quantum Monte Carlo Worm Algorithm  (WA) \cite{Worm}.
The superfluid stiffness in the $x$-direction was deduced from statistics of winding number, $W_x$,
fluctuations as $n_s = TL_x \langle W_x^2 \rangle$ \cite{Pollock1987}.
The instantaneous shape $h(x,\tau)$ of the edge at a given imaginary time $\tau$ is defined by summing up the number of particles $n_{x,y}(\tau)$ along $y$ for all grid points $x, \tau$ as
\be
h(x,\tau) =  \sum_{y=0}^{L_y-1} n_{x,y} (\tau) -1/2\;.
\label{ydef}
\ee
The equilibrium edge position at half filling is located at $h_0 = (L_y-1)/2$.
Since WA works in the Fock basis, the corresponding Monte Carlo estimator for $h(x,\tau)$
is based on straightforward processing of the many-body $\{ n_{i} (\tau) \}$ path-integral configuration.

\begin{figure}[t]
\includegraphics[width=1.0 \columnwidth]{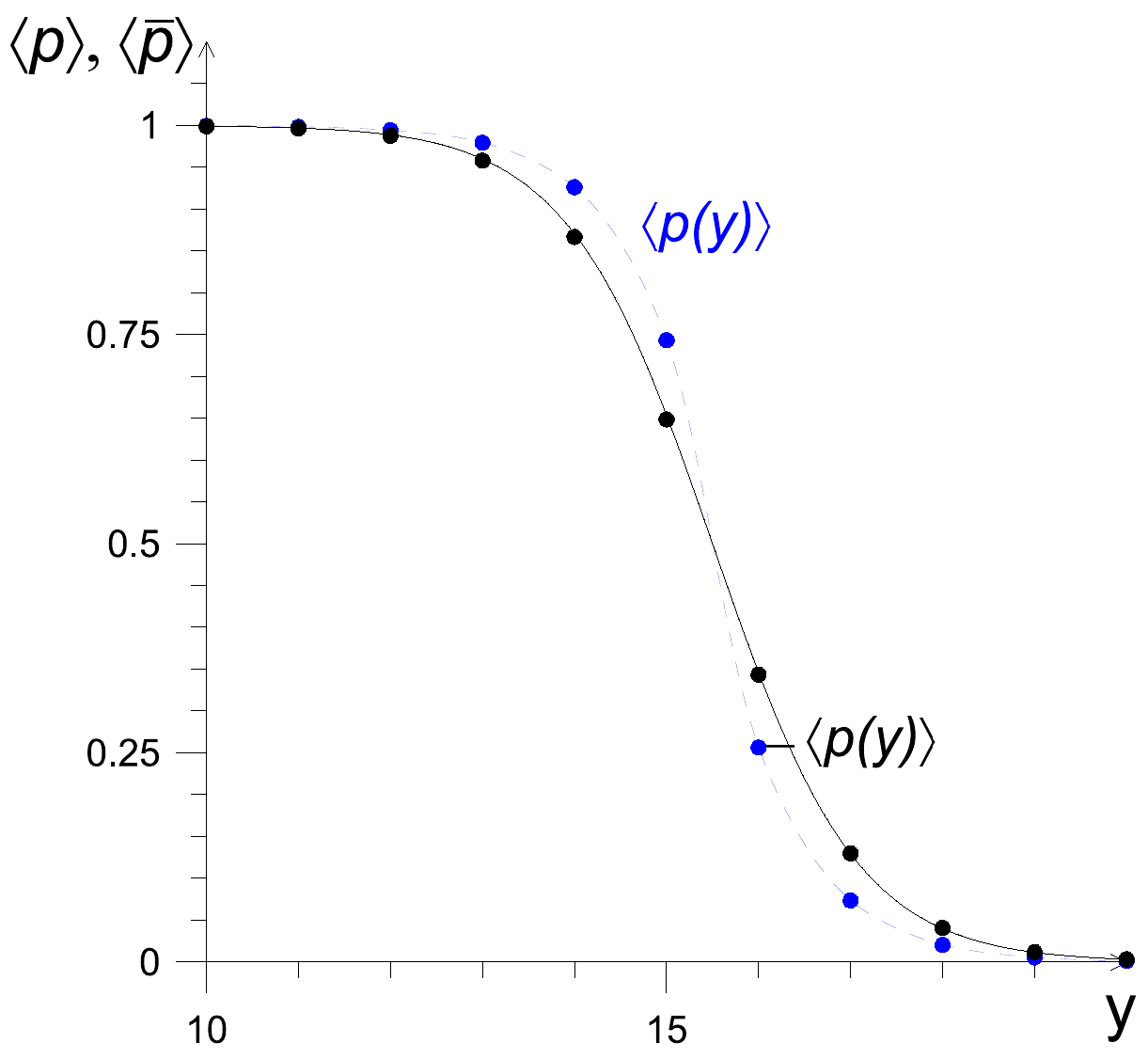}
\caption{Instantaneous and time-averaged edge profiles for model $(\ref{Hhc})$ at $V/t=-2.2$
and $\beta t =32$ for a system with $L_y=32$ and $L_x=64$. The pinning potential strength was $\delta \mu /t = 2$. Dashed line is to guide the eye. Solid line is the fit described in the text.
}
\label{profileV22}
\end{figure}

The density profiles across the edge (and its width) 
were obtained using density snapshots, $n_{x, y} (\tau)$, and time-averaged
density distributions $ \bar{n}_{x, y} = T\int_0^{\beta} n_{x,y}(\tau ) d\tau $. 
The profile coordinate was counted from the instantaneous
edge position
\be
p(x,y,\tau) = n_{x, y-h(x,\tau)+h_0} (\tau) \;.
\label{pdef}
\ee
The same protocol was applied to the time-averaged density distribution
\be
\bar{p}(x,y) = \bar{n}_{x, y-\bar{h}(x)+h_0} \;,
\label{pbardef}
\ee
with  $\bar{h}(x) =  T\int_0^{\beta} h(x,\tau ) d\tau $.
After statistical averaging both profiles---instantaneous and time-averaged---depend on $y$ coordinate only due to system's translation invariance in $x$ and $\tau$, and both functions are centered at the equilibrium position $h_0$.
In Fig.~\ref{profileV22}, we present these microscopic characteristics computed for $V/t=-2.2$. As expected,
the time-averaged profile is broader by accounting for dynamic fluctuations. By fitting the
$\langle \bar{p} \rangle$ profile to $\bar{p}(y)=[1+\tanh(2(y-h_0)/d)]/2$ we find
that for $V/t=-2.2$ the edge has a width of $d=3.15$. As we will see below,
this value is already large enough to guarantee 
that the Peierls potential for our system sizes 
is negligible and all low-energy/long-wavelength properties 
are governed by the TQF action.
%%%%%%%%%%%%%%%%%%%%%%%%%%%%%%%%%%%%%%%%%%%%%%%

\begin{figure}[H]
\includegraphics[width=1.05 \columnwidth]{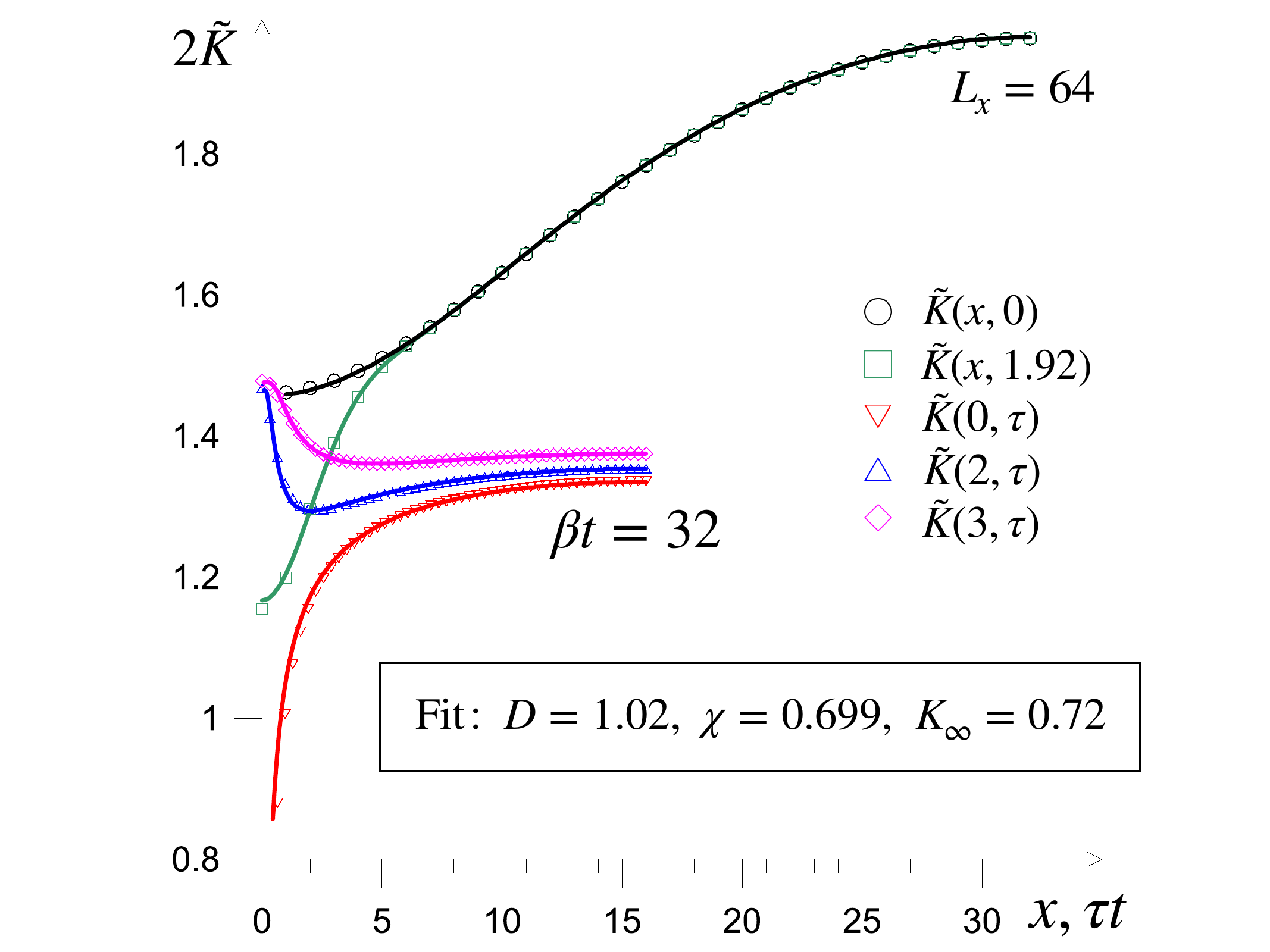}
\caption{Edge fluctuations in space and imaginary time for model $(\ref{Hhc})$
at $V/t=-2.2$ and $\beta t =32$ for a system with $L_y=32$ and $L_x=64$.
The data were fitted to the TQF function $\tilde K(x,\tau)$ (solid lines)
and the fit resulted in $D=1.02(2)$, $\chi=0.699(3)$, and $K_\infty=0.72(1)$.
}
\label{K2V22}
\end{figure}

Simulation data for $ 2\tilde{K}(x,\tau)  = \langle [ h(x,\tau) - h(0,0) ]^2  \rangle$
are presented in Fig.~\ref{K2V22} along with their fit using the equation
\be
\tilde{K}(x,\tau) = K_\infty+\frac{L}{2\beta \chi}\bar{\lambda}(r,t) ,
\label{fitK}
\ee
see Eqs.~(\ref{two_lambdas})--(\ref{lambda_1}).
Out of three fitting parameters ($K_\infty$, $\chi$, $D$), 
only $D$ is controlling the shapes of all curves in Fig.~\ref{K2V22}, while $K_\infty$ is responsible for their vertical shift and $\chi$ for the overall scale. The quality of asymptotic analytical predictions for domain wall shape fluctuations demonstrated by Fig.~\ref{K2V22} is remarkable: despite using only
large $x$ and $\tau$ points for the fit 
(the selection criterion was $x^2+(\tau t)^2 \ge 16$)
we observe near perfect agreement between the theory and simulations all the way to $x=1$ at $\tau=0$ and
$\tau t =0.5$ at $x=0$. This ``fingerprint" type of agreement leaves no doubt that we are dealing with the TQF system.

\begin{figure}[H]
\includegraphics[width=1.05 \columnwidth]{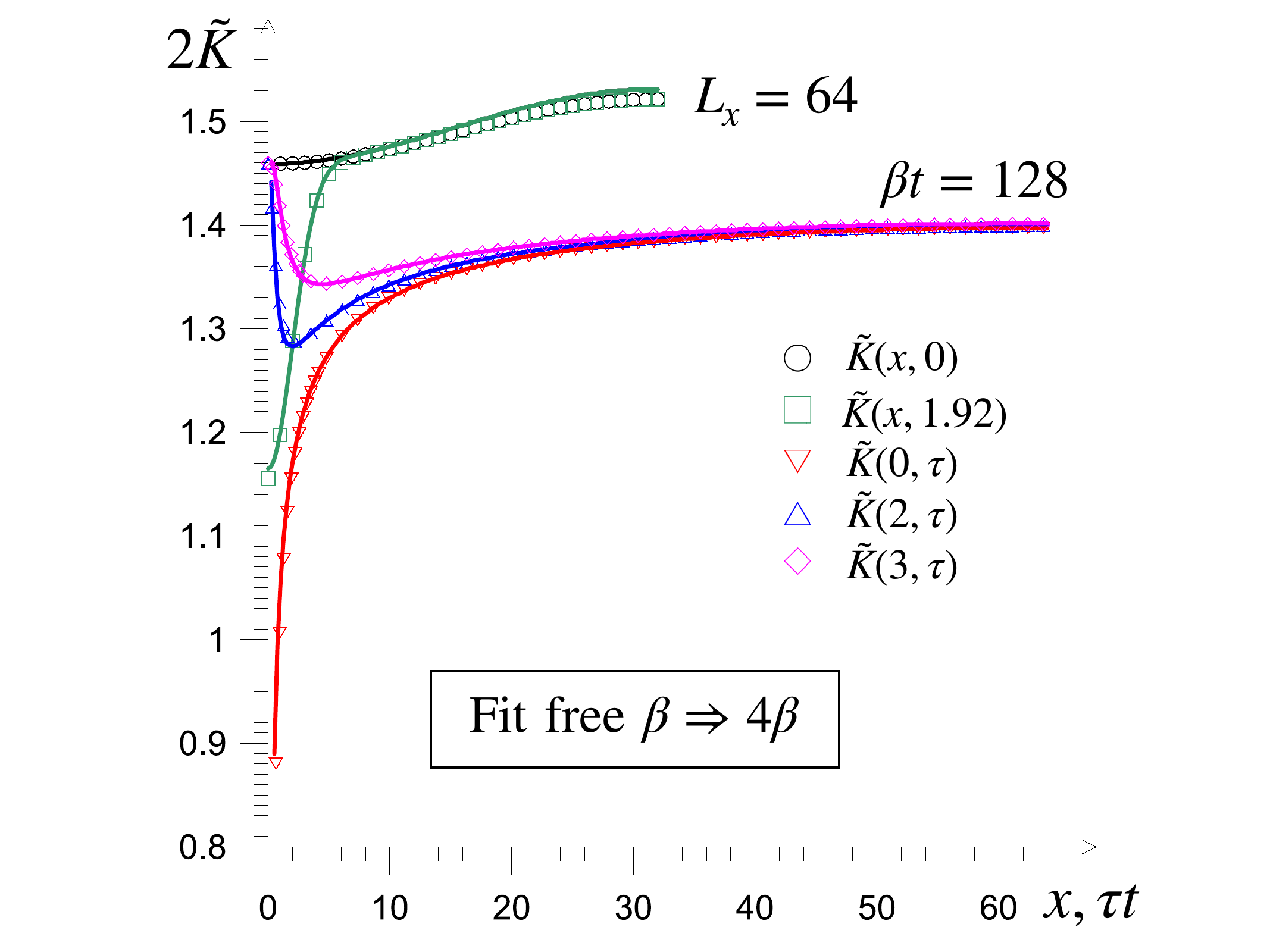}
\caption{Edge fluctuations in space and imaginary time for model $(\ref{Hhc})$
at $V/t=-2.2$ and $\beta t =128$ for a system with $L_y=32$ and $L_x=64$.
The data were fitted to the TQF function $\tilde K(x,\tau)$ (solid lines)
using parameters deduced from $\beta t=32$ simulations, see Fig.~\ref{K2V22}.}
\label{K2V22B128}
\end{figure}

The ultimate confirmation comes from agreement between the simulated superfluid stiffness,
$n_s = 1.492(2)$, and its value deduced from the $n_s=D^2/\chi = 1.49(4)$ relation. We are not aware of any other case where equilibrium system shape fluctuations would allow one to directly measure $n_s$. Finally, in Fig.~\ref{K2V22B128} we show a 
fit-free comparison between the simulation data and TQF predictions (keeping the same parameter set for $K_\infty$, $\kappa$, and $D$)  when going to a much lower temperature $\beta t =128$.
%%%%%%%%%%%%%%%%%%%%%%%%%%%%%%%%%%%%%%%%%%%%%%%
%%%%%%%%%%%%%%%%%%%%%%%%%%%%%%%%%%%%%%%%%%%%%%%
%%%%%%%%%%%%%%%%%%%%%%%%%%%%%%%%%%%%%%%

\begin{figure}[thb]
\includegraphics[width=1.05 \columnwidth]{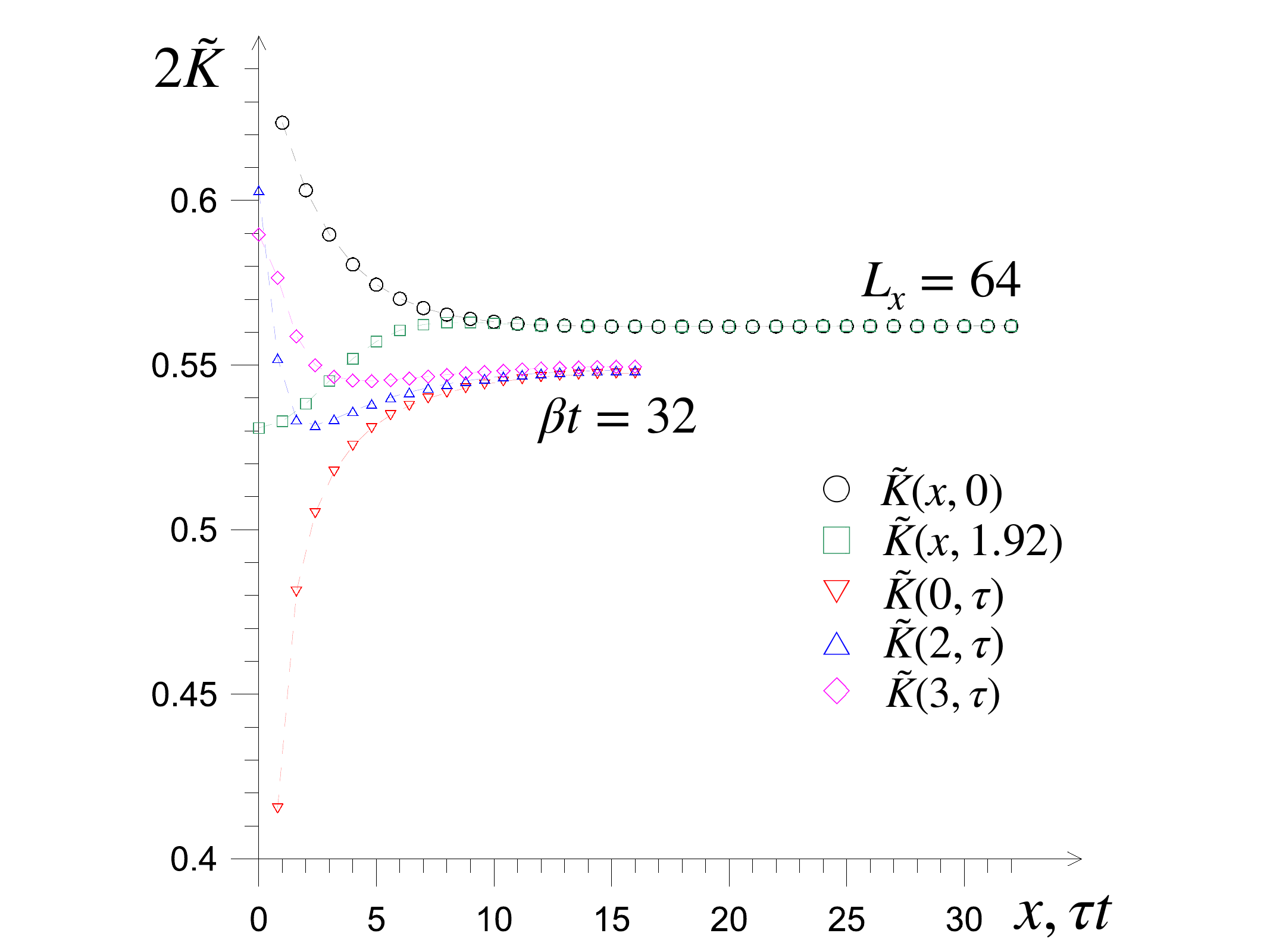}
\caption{Edge fluctuations in space and imaginary time for model $(\ref{Hhc})$
at $V/t=-2.5$ and $\beta t =32$ for a system with $L_y=32$ and $L_x=64$. Dashed lines are guides to the eye.
}
\label{K2V25}
\end{figure}

As the nearest-neighbor attraction is increased, the domain wall becomes narrower and, ultimately, the Peierls potential becomes relevant on length scales comparable or smaller than
the system size. This signals the expected crossover between the TQF and LL long-wave fluctuations \cite{Kuklov2024a}.
Figure~\ref{K2V25} presents simulation data for $\tilde{K}$ at $V/t=-2.5$ when the domain wall width
is about one lattice spacing, $d\approx 1.3$. These graphs cannot possibly be described by the TQF action because shape fluctuations in space are decreasing with distance, contrary to the TQF predictions (see Figs.~\ref{K2V22} and \ref{K2V22B128}), and quickly saturate to a constant value.
The domain wall of length $L_x=64$ is still displaying superfluid properties [with 
$n_s=0.59(1)$] but the LL parameter is already close to the critical value of $2$ for the 
standard superfluid-insulator BKT transition at integer filling factor. 
Any further increase of the attractive interaction results in an insulating wall. Similar behavior has been observed in model simulations of the  superclimbing dislocation \cite{MAX,LIU,Lode}.
%%%%%%%%%%%%%%%%%%%%%%%%%%%%%%%%%%%%%
%%%%%%%%%%%%%%%%%%%%%%%%%%%%%%%%%%%

\section{Two-component bosons}
\begin{figure}[H]
\includegraphics[width=0.9 \columnwidth]{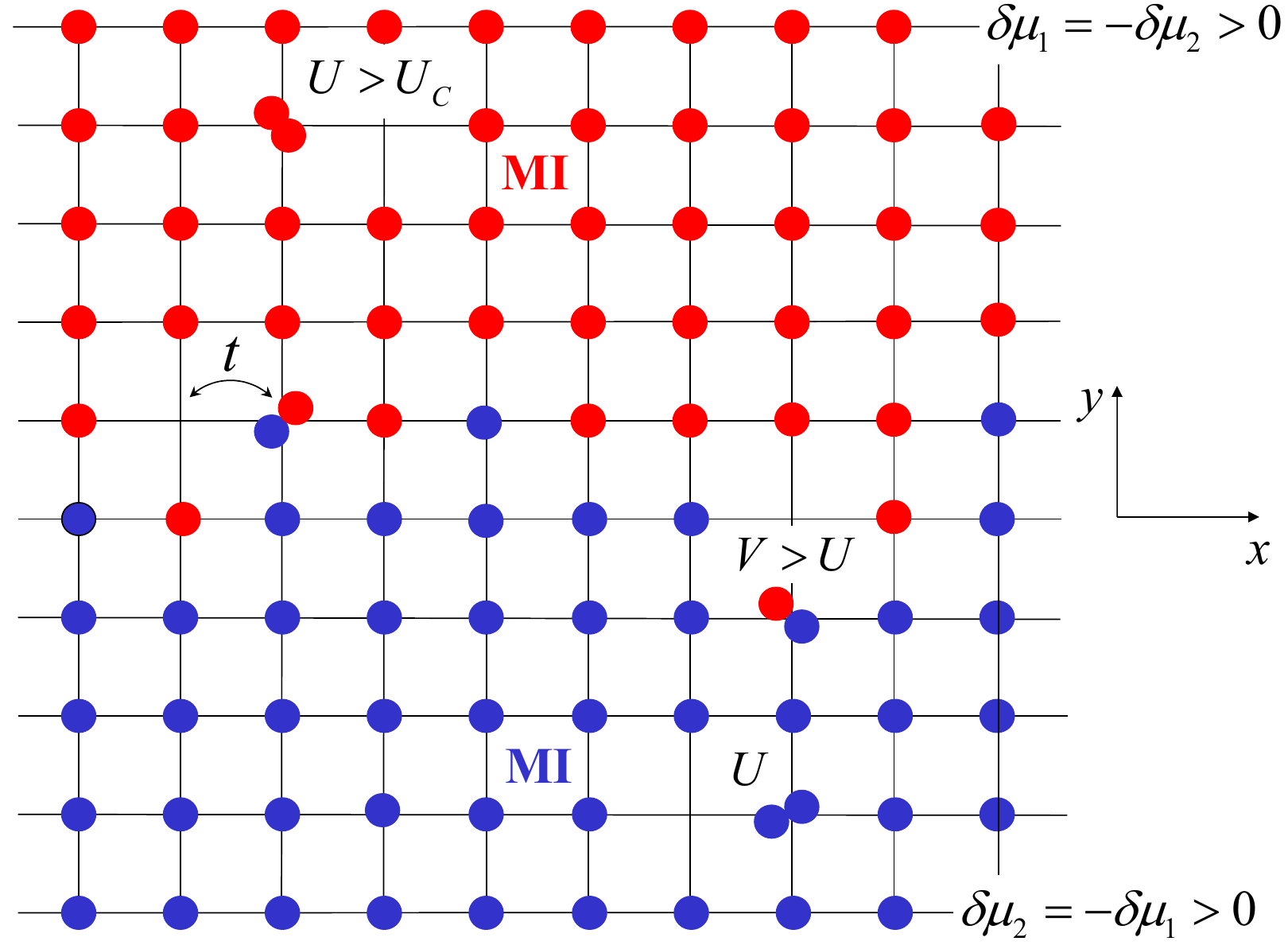}
\caption{ Two-component bosons (red and blue circles) at 
unity filling in the regime of phase separation. 
Counter-propagating motion by exchanging particles' places of different types along the boundary results in the
the boundary motion in the transverse direction. 
}
\label{fig:SCF}
\end{figure}

TQF physics is also expected to occur in the 2D lattice 
occupied by two species of repulsive bosons 
(labeled by index $\alpha=1,2$) at the total 
integer filling and in the regime of the phase separation between the components \cite{Kuklov2024a}. If the bulk of each spatially separated component is in the MI state, there is a regime when the boundary between the two species supports supercurrents with
counter-propagating flows of the components along the boundary,
i.e. it is an edge in the supercounterfluid state \cite{SCF}. 
An illustration of the system under discussion is shown in Fig.~\ref{fig:SCF}. 
The corresponding microscopic Hamiltonian is given by
\begin{eqnarray}
H&&= - t \sum_{\alpha=1,2; \langle i,j\rangle} b^\dagger_{\alpha, i} b_{\alpha, j} 
 \nonumber \\
&+& \sum_i\left[{U\over 2}(n^2_{1,i} +n^2_{2,i}) + V n_{1,i}n_{2,i}\right] ,   
\label{SCF}
\end{eqnarray}
where $b_{\alpha,i}$ are bosonic annihilation operators, and $n_{\alpha,i} =b^\dagger_{\alpha,i}b_{\alpha,i}$. 
[We consider the symmetric case when all parameters for different types of bosons are the same]. The onsite interaction constants $U>0, V>0$ are chosen in such a way that $U$ exceeds the critical value $U_c$ for MI phase in the single-component system, and $V>U$ ensures phase-separation when the two components are mixed. (If $V<U$ and $U>U_c$, the miscible state is in the counter-superfluid phase at low temperature \cite{SCF}.) 
The width of the supercounterfluid boundary between the two insulators is controlled by the proximity of $V$ to $U$  (at large enough $V$ the counter-transport ceases to exist). To establish a boundary oriented along the $x$-direction, it suffices to stabilize the structure by introducing potentials $\pm \delta \mu_1 =\mp \delta \mu_2$ at the lattice edges in the $y$-direction for these two species. The hopping across periodic boundary conditions (PBC) along the $y$-direction is turned off to ensure that the second boundary is sharp. If the other phase interface is thick enough, the Peielrs barrier restricting its transverse motion can be neglected and we obtain an edge in the TQF regime. 

%%%%%%%%%%%%%%%%%%%%%%%%%%%%%%%%%%%%%
\begin{figure}[thb]
\includegraphics[width=0.8 \columnwidth]{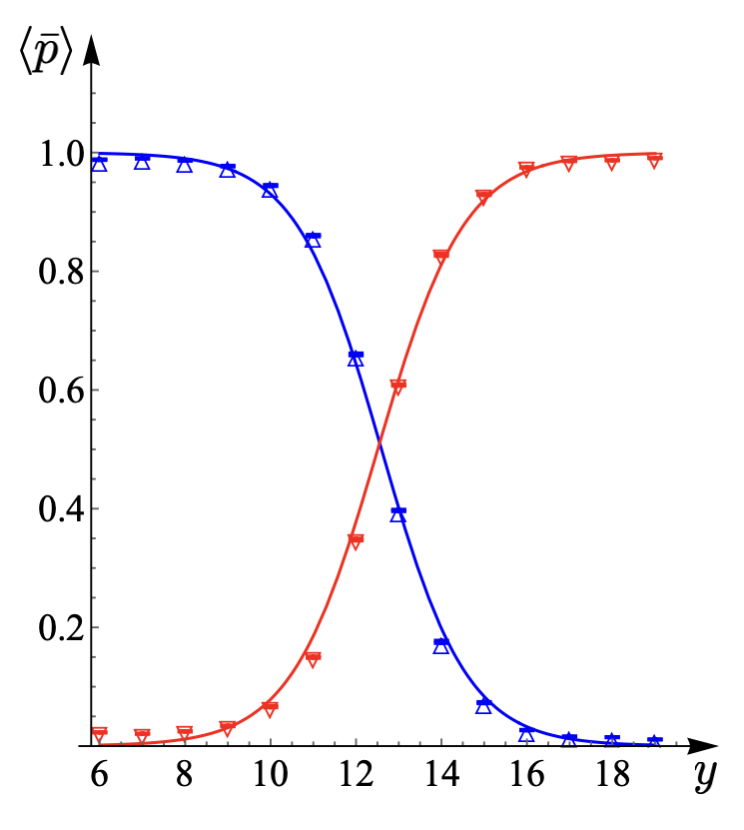}
\caption{Time-averaged edge profiles of two species (red and blue) for model (\ref{SCF})
at $U/t=18$, $V/t=19$, $\delta \mu_1 =\delta \mu_2=t$  and $\beta t =16$ for a system with $L_y=24$ and $L_x=16$. Solid lines are the fit described in the text.}
\label{twoprofile}
\end{figure}
%%%%%%%%%%%%%%%%%%%%%%%%%%%%%%%%%%%

The instantaneous shape, $h(x,\tau)$, of the edge at a given imaginary time $\tau$, is defined by summing up the number of particles of the first species, $n_{1, x,y}(\tau)$, along $y$ for all grid points $x, \tau$, but only if they reside on the side predominantly occupied by the second species; to be precise, 
$\alpha =1$ bosons are in ``majority" at $L_y/2+1$ to $L_y$, while $\alpha =2$ bosons are in ``majority" at $1$ to $L_y/2$, see Fig.~\ref{fig:SCF}. 
\be
h(x,\tau) =  \sum_{y=1}^{L_y/2} n_{1, x, y } (\tau) -  \sum_{L_y/2+1}^{L_y} n_{2, x, y } (\tau)  \;.
\label{2ydef}
\ee
Since there is no difference between these two species, the equilibrium edge position at half filling is located at $h_0 = (L_y+1)/2$.
The density profiles across the edge (and its width) were obtained using time-averaged density distributions $ \bar{n}_{\alpha, x, y} = T\int_0^{\beta} n_{\alpha, x, y}(\tau ) d\tau $. The rest of the numerical protocol for obtaining profiles is identical to what was described above for the hard-core case.  In Fig.~\ref{twoprofile}, we present this microscopic characteristic   computed for $U/t=18.0$, $V/t=19.0$ and $\delta \mu_1=\delta \mu_2=t$. By fitting both profiles to $\bar{p}(y)=[1 \pm \tanh(2(y-h_0)/d)]/2$ we find that for this parameter set the edge has a width of $d=4.05(7)$. We expect---and the simulations confirm---that for this width the Peierls potential for our system size and temperature is negligible.
%%%%%%%%%%%%%%%%%%%%%%%%%%%%%%%%%%%%%
\begin{figure}[thb]
\includegraphics[width=0.85 \columnwidth]{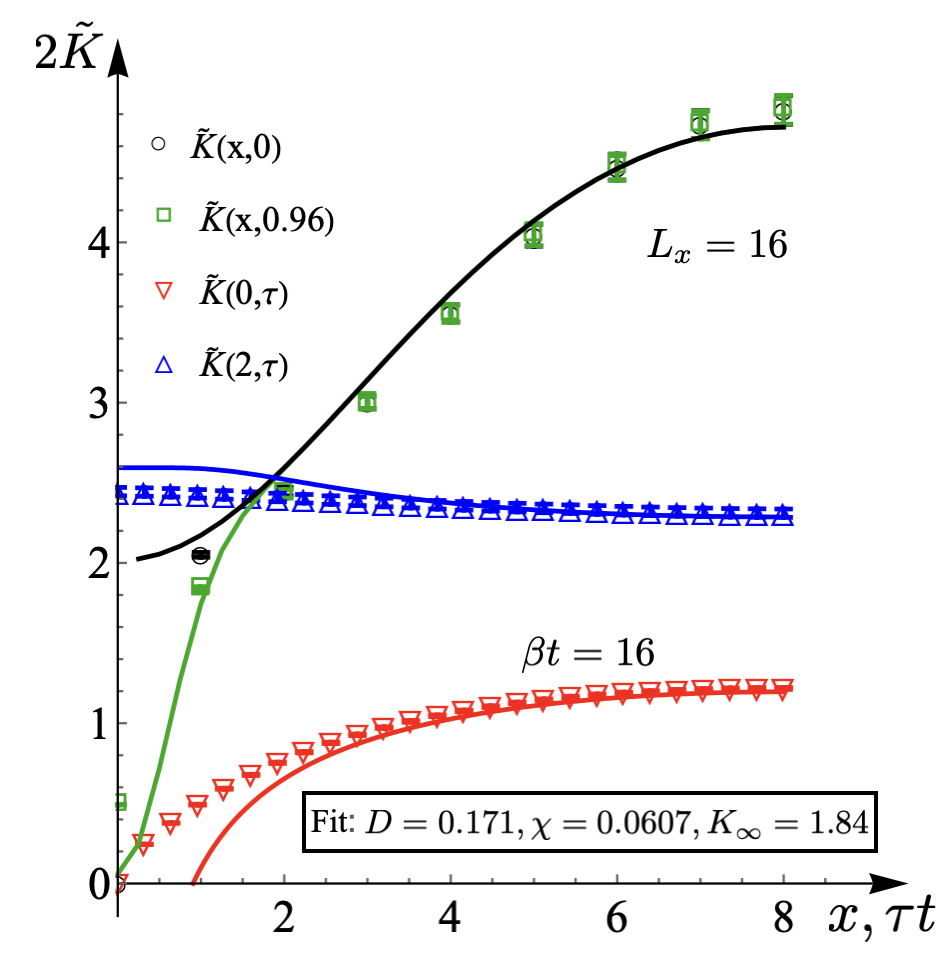}
\caption{Edge fluctuations in space and imaginary time for model (\ref{SCF})
at $U/t=18$, $V/t=19$, $\delta \mu_1 =\delta \mu_2=t$  and $\beta t =16$ for a system with $L_y=24$ and $L_x=16$. 
}
\label{twocomponent}
\end{figure}
%%%%%%%%%%%%%%%%%%%%%%%%%%%%%%%%%%%

In Fig.~\ref{twocomponent}, we compare TQF predictions with simulation data for model (\ref{SCF}). Since insulating bulk  states are close to the quantum critical point, Monte Carlo simulations with two worms are far more demanding than for the single-components hard-core system, explaining why we limit ourselves here with system size $L_y=24$, $L_x=16$.  
The validity of asymptotic analytical predictions for domain wall shape fluctuations, as demonstrated by Fig.~\ref{twocomponent}, is evident. Despite only using large $x$ and $\tau$ points for the fit (with the selection criterion being $x^2+(\tau t)^2 \ge 16$), good agreement between the theoretical framework and simulation outcomes extends to smaller values of $x$ and $\tau t  \sim 3$; deviations at $|x|, \tau t < 2$ are expected because the  edge width is large. 
The TQF interpretation gains final support from the agreement between the simulated supercounterfluid stiffness, $n_s = 0.51(4)$, and its value deduced from the relation $n_s=D^2/\chi = 0.48(3)$.

%%%%%%%%%%%%%%%%%%%%%%%%%%%%%%%%%%%

\section{Incomplete helium-4 layer on graphite}

In order to identify a real-material system in which some of the above  predictions could be tested experimentally, we have carried out microscopic numerical simulations at low temperature of an incomplete monolayer of $^4$He adsorbed on a graphite substrate, and studied the behavior of the system near the free edge, where a TQF may exist. 

Thin films of $^4$He on graphite have been extensively investigated experimentally, mainly because the strong attractiveness of this substrate, and its pronounced corrugation, lead to the stabilization of crystalline phases of $^4$He not observed in the bulk. (On weakly attractive substrates $^4$He does not crystallize at low temperature; rather, ``wetting,'' i.e., continuous growth of a superfluid film as a function of chemical potential is observed \cite{Boninsegni1999,VanCleve2008}.) It is known both experimentally and theoretically \cite{Nielsen,Corboz2008} that the equilibrium phase of a $^4$He monolayer on graphite is a commensurate crystal, registered with the underlying carbon substrate. Such a phase, which corresponds to a coverage (effective 2D density) $\theta_0=0.0636$ \Am2, is commonly referred to as $C_{1/3}$, as \he4 atoms occupy one of the three equivalent sublattices of preferred adsorption sites. As the coverage is increased, a transition to an incommensurate crystalline layer takes place, while for coverages below $\theta_0$ coexistence of solid regions of coverage $\theta_0$ and low-density vapor is predicted \cite{Pierce1999}. This is therefore a well-suited physical system in which the physical picture described above can be investigated experimentally.

%%%%%%%%%%%%%%%%%%%%%%%%%%%%%%%%%%%%%
\begin{figure}[thb]
\includegraphics[width=1. \columnwidth]{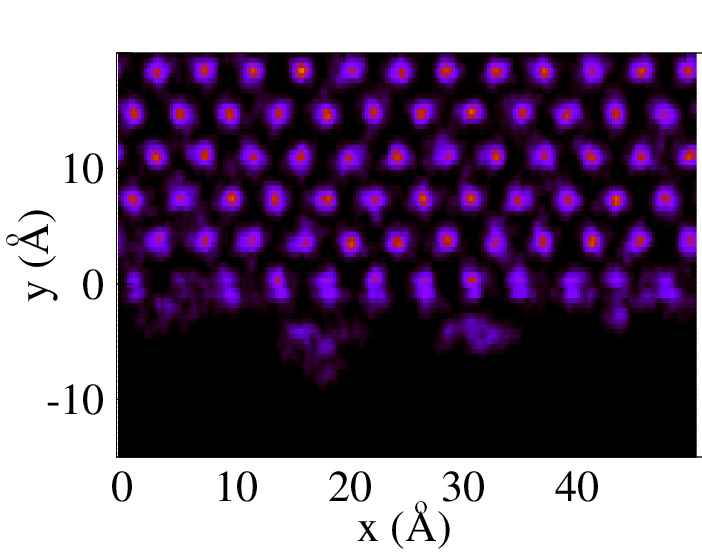}
\caption{Representative density map (particle world lines) for an incomplete \he4 monolayer adsorbed on graphite, a temperature $T=0.25$ K. Only the region in the vicinity of the free edge is shown. The total number of \he4 is 144. The crystalline phase has 2D density $\theta_0$ (see text).
}
\label{he4edge}
\end{figure}
%%%%%%%%%%%%%%%%%%%%%%%%%%%%%%%%%%%

We studied by computer simulation an incomplete commensurate crystalline $^4$He monolayer on graphite in thermodynamic equilibrium at low temperature, by making use of the same microscopic model and computational methodology utilized in Ref. \cite{Corboz2008}, with the only difference that we carried out canonical (i.e., constant density) simulations.  
Figure~\ref{he4edge} shows a representative snapshot (2D density obtained from particle world lines) of the region near the free edge of an incomplete \he4 monolayer adsorbed on graphite (the system comprises altogether 144 \he4 atoms), at a temperature $T$=0.25 K. The position of the edge (bottom row of atoms), classically, is at $y=0$.

%%%%%%%%%%%%%%%%%%%%%%%%%%%%%%%%%%%%%
\begin{figure}[thb]
\includegraphics[width=1. \columnwidth]{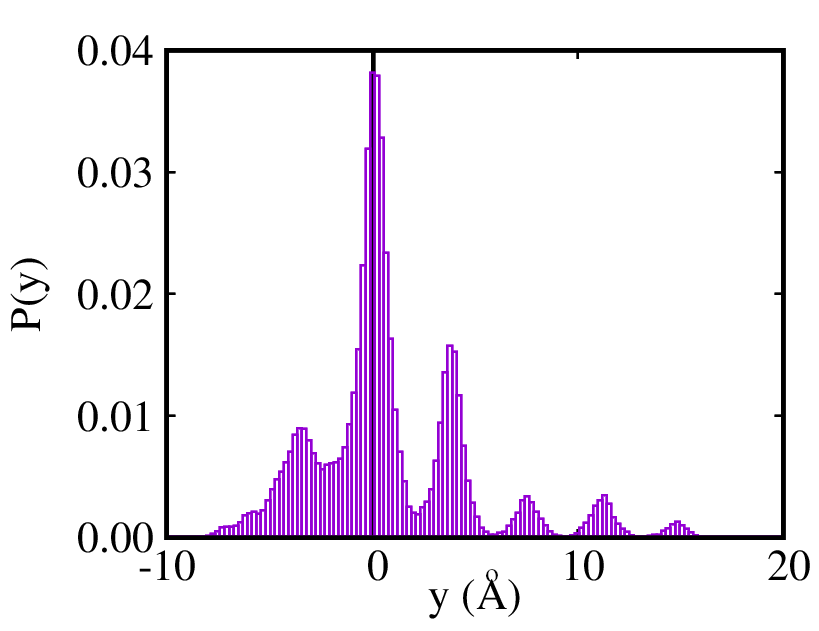}
\caption{Computed (unnormalized) probability for a \he4 atom to be part of a cycle of exchange, as a function of its position $y$ in the direction perpendicular to a free edge. In this case, the crystalline sample occupies (classically) the $y > 0$ region.
}
\label{frequency}
\end{figure}
%%%%%%%%%%%%%%%%%%%%%%%%%%%%%%%%%%%

The most important observation is that, while the underlying crystalline monolayer (of 2D density $\theta_0$) remains stable away from the free edge, atoms near the edge are significantly delocalized and can ``climb'' on top of their nearest neighbor, i.e., the interface is roughened by quantum fluctuations. This leads to a local enhancement of quantum-mechanical exchanges of \he4 atoms, which are essentially non-existent away from the edge (or, in a complete monolayer). 

This is quantitatively illustrated in Fig.~\ref{frequency}, which displays the computed probability for a \he4 atom to be part of a cycle of exchange, as a function of its position along the direction perpendicular to the free edge. The crystalline layer occupies the $y > 0$ region, but \he4 atoms in the vicinity of the edge are allowed to wander away from it (i.e., into the $y <0$ region), creating instantaneous vacancies, which in turn can be filled by other atoms, ultimately leading to cycles of exchanges. These exchanges are inhibited in the crystal by the short-distance hard-core repulsion of the helium pairwise interaction. We see that, at the temperature of this particular simulation, exchanges extend up to approximately four layers into the crystal.

All of this constitutes a strong indication that at sufficiently low temperature a TQF may exist in this system, near the free edge. In a simulation like the one described here, superfluidity is connected to the winding around the periodic boundary conditions (in the direction along the edge) of the many-particle world lines \cite{Pollock1987}. Obviously, simulations carried out on systems of sufficiently large size are required in order to establish this conclusion robustly, unambiguously detecting any signal coming from the region near the edge (as opposed to spurious effects arising from the finite width of the sample, in the direction perpendicular to the edge). Such a  comprehensive study is presently in progress. It is also worth investigating the same system with reduced particle mass and/or substrate potential to increase quantum delocalization effects, not to mention that incomplete layers of $^3$He particles is yet another potentially 
interesting setup. 

%%%%%%%%%%%%%%%%%%%%%
\section{Conclusion}
%%%%%%%%%%%%%%%%%%%%%

Motivated by recent theoretical progress, we performed numerically exact simulations of characteristic microscopic models featuring transverse quantum fluid (TQF) states in the edges/interfaces. The key feature distinguishing TQF from an incoherent transverse quantum fluid (iTQF) are the climbing degrees of freedom canonically conjugate to the field of superfluid phase and responsible for the formation of superclimbing normal modes described by the Hamiltonian (\ref{H_TQF}). Our goal was to check theoretical prediction that quantum fluctuations of the superclimbing modes control long-wave correlations of the edge/interface height. A delicate aspect of this prediction is that it is supposed to work under the condition of microscopic quantum roughness, while, in accordance with the theory itself, Peierls barrier eventually becomes relevant in the asymptotic long-wave/low-temperature limit and transforms TQF into a LL
with exponentially large LL parameter.
Yet another subtle aspect is related to the predicted properties of the equal-time correlator of the universal quantum fluctuations of the edge/interface height---the most natural direct observable in both experiment and simulations. The correlator is a featureless constant in the zero-temperature limit, meaning that  one has to use a low but finite temperature as a resource for resolving the universal quantum character of the long-wave 
equal-time correlations of the height.
The same is also true for correlations of the superfluid phase
field, since the two fields are described by the same [up to exchanging parameters $\chi$ and $n_s$ places] effective action. 

In light of these subtleties, the first question our simulations were supposed to clarify was about the existence of a reasonably large region in the space of model parameters, including the range of finite temperatures and system sizes, where the desired universal physics would be clearly observed. Our numeric results, demonstrating impressive agreement with analytic predictions---even at unexpectedly short distances  on the order of few lattice spacings---clearly demonstrate that such a region does exist. 

Quantitatively, the observed fingerprint universal features clearly distinguish the TQF state not only from the standard LL but also from the cousin iTQF state. At the qualitative level, we numerically demonstrated that quantum fluctuations of the edge/interface height are controlled by---and thus allow one to extract---the superfluid stiffness. This is the remarkable manifestation of the crucial circumstance behind the superclimbing modes: the fields of the height and superfluid phase are canonically conjugate to each other.

%{\color{red} An interesting aspect of these models, which makes them distinct from the superclimbing dislocation \cite{sclimb,Kuklov2022,Radzihovsky2023} and the iTQF \cite{Kuklov2024b}, is the interplay between the criticality determining the divergence of the width of the domain wall (as $\chi \to 0$ on the approach to the critical point in the bulk) and the TQF fluctuations---especially in the context of different $z$ exponents ($z=1$ for the criticality and $z=2$ for the TQF). At the moment this question remains open.}

\bigskip

\begin{acknowledgements}
AK, BS, and NP acknowledge support from the National Science Foundation under Grants DMR-2335905 and DMR-2335904. MB acknowledges support from the Natural Science and Engineering Research Council of Canada, under grant RGPIN 2018-04227. CZ acknowledges support from the National Natural Science Foundation of China (NSFC) under Grants 12204173 and 12275002.
\end{acknowledgements}

%%%%%%%%%%%%%%%%%%%%%
\end{document}